\documentclass[prd,aps,preprint,tightenlines,nofootinbib,superscriptaddress]{revtex4}
\usepackage{epsfig}
\usepackage{amsmath}
\input{epsf}
\usepackage{psfrag}

\usepackage[usenames,dvipsnames]{color}
\usepackage[normalem]{ulem} 

\begin{document}


\vspace*{2cm}
\title{Exploring gluon tomography with polarization dependent diffractive J/$\psi$ production }

\author{ James~Daniel~Brandenburg}
 \affiliation{\normalsize\it Brookhaven National Laboratory, New York, USA }
 
 \author{Zhangbu~Xu }
 \affiliation{\normalsize\it  Brookhaven National Laboratory, New York, USA }
 
  \author{Wangmei~Zha }
 \affiliation{\normalsize\it University of Science and Technology of China, Hefei, China }

\author{Cheng~Zhang}
\affiliation{\normalsize\it Department of Physics, Center for Field Theory and Particle Physics, Fudan University, Shanghai, 200433, China}
\affiliation{\normalsize\it Key Laboratory of
Particle Physics and Particle Irradiation (MOE),Institute of
Frontier and Interdisciplinary Science, Shandong University,
QingDao, China }

\author{Jian~Zhou}
 \affiliation{\normalsize\it Key Laboratory of
Particle Physics and Particle Irradiation (MOE),Institute of
Frontier and Interdisciplinary Science, Shandong University,
QingDao, China }

\author{Yajin Zhou}
\affiliation{\normalsize\it Key Laboratory of Particle Physics and
Particle Irradiation (MOE),Institute of Frontier and
Interdisciplinary Science, Shandong University, QingDao,  China }

\begin{abstract}
We study azimuthal asymmetries in diffractive J/$\psi$ production  in ultraperipheral heavy-ion collisions at RHIC and LHC energies using the color glass condensate effective theory. Our calculation successfully describes azimuthal averaged $J/\psi$ production cross section measured by STAR and ALICE.  We further predict very large $\cos 2\phi$ and $\cos 4\phi$ azimuthal asymmetries for diffractive $J/\psi$ production both in UPCs at RHIC and LHC energies and in eA collisions at EIC energy. These novel polarization dependent observables may provide complementary information for constraining gluon transverse spatial distribution inside large nuclei. As compared to all previous analysis of diffractive $J/\psi$ production, the essential new elements integrated in our theoretical calculations are: the double-slit interference effect, the linear polarization of coherent photons, and the final state soft photon radiation effect. 
\end{abstract}

\maketitle

\section{Introduction} 
Diffractive processes in high-energy scatterings have long been considered as a powerful tool to study saturation physics, to explore the multi-dimensional structure of gluonic matter inside nuclei/nucleons, and to resolve the mass structure of the proton~\cite{Kharzeev:1995ij,Hatta:2019lxo,Kou:2021bez,Guo:2021ibg,Sun:2021gmi,Sun:2021pyw}. An especially interesting diffractive process is the exclusive production of vector mesons in collisions between a real or virtual photon with a target that remains intact after scattering.  At relatively large $x$, such an exclusive process is formulated in terms of the square of the gluon PDF in the leading log approximation~\cite{Ryskin:1992ui}, or more rigorously in terms of generalized parton distributions within collinear factorization~\cite{Ji:1998pc,Vanttinen:1998en,Ivanov:2004vd,Koempel:2011rc}. At small $x$ it can be described using the dipole model~\cite{Brodsky:1994kf} or the color glass condensate effective theory to incorporate multiple gluon rescattering effects.  Along these research lines, tremendous theoretical efforts~\cite{Klein:1999qj,Munier:2001nr,Kowalski:2003hm,Kowalski:2006hc,Strikman:2008zz,Lappi:2010dd,Rebyakova:2011vf,Guzey:2013qza,Guzey:2013jaa,Xie:2015gdj,Guzey:2016piu,Guzey:2016qwo,Yu:2017pot,GayDucati:2017ksh,Luszczak:2019vdc,Lansberg:2018fsy,Cai:2020exu,Henkels:2020qvo,Lappi:2020ufv,Guzey:2020ntc,Yu:2021sxg,Mantysaari:2021ryb,Mantysaari:2020lhf,Eskola:2022vpi,Frankfurt:2022jns,Ma:2022rfl} haven been made to understand the underlying physics of this process in the past three decades.

Among various exclusive vector meson production processes, one important channel is the production of the $J/\psi$ meson. On the one hand, the charm quark is sufficient heavy to justify perturbative treatment, while on the other its mass is not too large to allow access to the saturation regime.  In addition, $J/\psi$ can be relatively easily identified experimentally with sizable production cross section, for example in ep collisions at HERA~\cite{ZEUS:2002wfj,ZEUS:2004yeh,H1:2013okq}. $J/\psi$ also can be exclusively produced in ultraperipheral heavy-ion collisions where one of the colliding nuclei serves as a quasi-real photon source while the other plays the role of a target.  In recent years, there have been many active experimental programs devoted to studying this process at the RHIC and the LHC~\cite{PHENIX:2009xtn,Contreras:2013oan,STAR:2021wwq,Schmidke:2016ccw,CMS:2016itn,Khachatryan:2016qhq,STAR:2019yox,ALICE:2019tqa,ALICE:2021gpt,ALICE:2021tyx,LHCb:2021bfl}. The main benefit provided by UPCs is the extremely-high luminosity of quasi-real photons, which renders to us the opportunity to study the exclusive $J/\psi$ production process with very high precision.  Exclusive $J/\psi$ production  can also be studied at the future EIC and EicC. Some interesting discussions on coherent $J/\psi$ production at EIC can be found in Refs.~\cite{Lomnitz:2018juf,Mantysaari:2020lhf}.

In the present work, we will address this topic from a different angle, namely investigating the azimuthal dependence of the coherent $J/\psi$ production cross section in UPCs and in eA collisions. Recently, a large $\cos 2\phi$ and $\cos 4\phi$ azimuthal asymmetries in diffractive $\rho^0$ production in UPCs has been reported by the STAR collaboration~\cite{daniel:2019,STAR:2022wfe}, where $\phi$ is the azimuthal angle between the $\rho^0$ transverse momentum and the transverse momentum carried by its decay product pion particles.  It has been found in Refs.~\cite{Xing:2020hwh,Zha:2020cst} that the $\cos 2\phi$ azimuthal asymmetry essentially arises from the linear polarization of the incident coherent photons. Such phenomenon was not recognized until very recently~\cite{Li:2019yzy,Li:2019sin,Xiao:2020ddm,Wang:2022gkd}, and was quickly verified by the STAR collaboration~\cite{Adam:2019mby} via the measurement of the characteristic $\cos 4\phi$ azimuthal asymmetry in purely electromagnetic dilepton production in UPCs. 
By coupling with the elliptic gluon distribution, the linearly polarized coherent photon distribution also plays an important role in inducing the $\cos 4\phi$ asymmetry in exclusive $\pi^+$ $\pi^-$ pair production in UPCs~\cite{Hagiwara:2021qev}. Moreover, sizable $\cos \phi$ and $\cos 3\phi$ azimuthal asymmetries that result from the Coulomb nuclear interference effect have been predicated in Ref.~\cite{Hagiwara:2020juc}. All in all, the linear polarization of coherent photons has been proven to be a powerful experimental tool for exploring novel QCD phenomenology as well as novel aspects of QED under extreme conditions~\cite{Baur:1998ay,Klein:2020fmr,Steinberg:2021lfm,Hattori:2020htm,Copinger:2020nyx,Brandenburg:2021lnj}. 

These polarization dependent observables are very sensitive to nuclear geometry and thus provide a complementary way to extract transverse spatial gluon distribution. In this paper, we  investigate the azimuthal dependent production of $J/\psi$, a case for which the mass of the charm quark sets a hard scale, justifying a perturbative treatment.  A similar azimuthal asymmetry arising from the linear polarized photon distribution is also expected in $J/\psi$ production. Apart from this source, the final state soft photon emitted from the produced dilepton pair can also give rise to significant azimuthal asymmetries due to the mechanism discovered in Refs.~\cite{Catani:2014qha,Catani:2017tuc,Hatta:2020bgy,Hatta:2021jcd}.  In our calculation, we will also employ an impact parameter dependent formalism~\cite{Xing:2020hwh} to naturally incorporate the double slit interference effect~\cite{Klein:1999gv,Abelev:2008ew,Zha:2018jin,Xing:2020hwh,Zha:2020cst}. We will demonstrate below that in order to correctly account for the absolute normalization of the cross section, as well as for the $t$-dependence of, both the azimuthal averaged cross section and the $\cos 2\phi$ asymmetry, it is crucial to simultaneously take into account the double-slit interference effect, the linear polarization of the coherent photons, and the final-state soft-photon radiation effect -- all of which were overlooked in previous analysis of diffractive $J/\psi$ production in UPCs\footnote{Shortly after our paper being submitted to Arxiv, there appeared another calculation of the diffractive $J/\psi$ production in UPCs where the interference effect has been included in their analysis as well~\cite{Mantysaari:2022sux}.}.
 
The paper is organized as follows. In Sec.II, we derive cross section formulas with all order soft photon radiation resummation being performed. In Sec.III, we present the numerical estimations of the azimuthal averaged $J\psi$ production cross section and compare it with the experimental measurements. We further make predictions for the azimuthal asymmetries in exclusive $J/\psi$ production in UPCs at RHIC and LHC energies, and that in eA collisions for EIC kinematics. Finally, the paper is summarized in Sec. IV.

 \section{Theoretical set up}
 In this section, we derive the azimuthal dependent cross section of exclusive di-lepton production in UPCs via $J/\psi$ decay.  First, we briefly review the calculation of the exclusive $J/\psi$ production amplitude in the dipole model.  In the dipole model it is a common practice to divide vector meson photoproduction process into three steps: quasi-real photon splitting into a quark and anti-quark pair, the color dipole scattering off a nucleus, and the subsequently recombining to form a vector meson after penetrating the nucleus target. Following this picture, it is straightforward to write down the scattering amplitude for both coherent and incoherent production,
 ${\cal A}_{co}(\Delta_\perp)$ and ${\cal A}_{in}(\Delta_\perp)$ which are given by,
\begin{eqnarray}
 {\cal A}_{co}(x_g,\Delta_\perp) \!\!&=&\!\! \int \!d^2
b_\perp e^{-i \Delta_\perp \cdot b_\perp} \!\int \frac{d^2
r_\perp}{4\pi}  \ N(r_\perp,b_\perp) [\Phi^*\!K](r_\perp)
 \nonumber \\
 {\cal A}_{in}(x_g,\Delta_\perp) \!\!&=&\!\! \sqrt{\!A }2 \pi B_p e^{-\!B_p\Delta_{\! \perp}^2\!/2}
    \left [  \int \!\frac{d^2
r_\perp}{4\pi}  {\cal N}(r_\perp)e^{\!-2\pi (A\!-\!1)B_p T_A(b_\perp) {\cal N}(r_\perp) }
[\Phi^*\!K](r_\perp) \right ] \ \ \
\end{eqnarray}
where ${\cal N}(r_\perp)$ is the dipole-nucleon scattering amplitude. $N(r_\perp,b_\perp)$ is the elementary amplitude for the scattering of a dipole of size $r_\perp$ on a target nucleus at the impact parameter $b_\perp$ of the photon-nucleus collision. $T_A(b_\perp)$ is the nuclear thickness function and $[\Phi^*\!K]$ denotes the overlap of the photon wave function and the vector meson wave function,
\begin{eqnarray}
[\Phi^*\!K](r_{\! \perp})\! =\!\frac{N_{\!c} e e_q}{\pi}\!\! \int_0^1 \!\!\! dz\!
\left \{m_q^2 \Phi^*(|r_\perp|,z)K_0(|r_\perp| e_f) \!+\!
\left [  z^2\!+\!(1\!-\!z)^2   \right ]
 \!  \frac{\partial\Phi^*(|r_\perp|,z)}{\partial |r_\perp|}
 \frac{\partial  K_0(|r_\perp| e_f)}{\partial |r_\perp|} \!\right \}
\end{eqnarray}
where $z$ stands for the fraction of the photon's light-cone momentum carried by the quark, and $e_f \approx m_q$. $\Phi^*(|r_\perp|,z)$ is the scalar part of the vector meson wave function. Here we ignore a phase arising from the non-forward effect~\cite{Hatta:2017cte,Hagiwara:2020mqb}.

One can easily derive the dilepton production amplitude  by multiplying the $J/\psi$ production amplitude with a simplified Breit-Wigner form which describes the transition from $J/\psi$ into $e^+ \ e^-$,
\begin{eqnarray}
{\cal M}_{J/\psi \rightarrow e^+ e^-}= i\left [ {\cal A}_{co}(x_g,\Delta_\perp)+{\cal A}_{in}(x_g,\Delta_\perp) \right ] 
 \frac{  \hat k_\perp^{\mu}\bar u(p_1) \gamma_\mu v(p_2)}{Q^2-M^2+iM \Gamma} \frac{-2e^2e_q\delta^{ij}}{M\sqrt {M}}\phi(0).
\end{eqnarray}
 Here $M$ denotes the $J/\psi$'s  mass, $\hat k_\perp $ is the incident coherent photon's polarization vector which is parallel to it's transverse momentum, $Q$ is the invariant mass of the dilepton system, and $P_\perp$ is defined as $P_\perp=(p_{1\perp}-p_{2\perp})/2$ with $p_{1\perp}$ and $p_{2\perp}$ being the produced lepton's transverse momenta. $\phi(0)$ is the wave function for the charm quark inside $J/\psi$ at the origin. The decay width of $J/\psi$ from this channel is related to the zero point wave function through $\Gamma=16 \pi \alpha_e^2 e_q^2 \frac{|\phi(0)|^2}{M^2} $.

For the photoproduction of vector mesons in UPCs, it is important to take into account the double-slit-like quantum interference effect~\cite{Klein:1999gv,Abelev:2008ew,Zha:2018jin}.  To this end, we developed a joint impact parameter dependent and $q_\perp$ dependent cross section formula ~\cite{Xing:2020hwh}, in which the double-slit interference effect is naturally included.  Such a  formalism has been employed to compute the azimuthal 
dependent cross section for diffractive $\rho^0$ photoproduction in UPCs~\cite{Xing:2020hwh,Hagiwara:2021qev}. It has been found that the $t$ dependence of the cross section is significantly modified by this interference effect, particularly at mid rapidity.  Following the method outlined in Ref.~\cite{Xing:2020hwh}, we derive the impact parameter dependent differential cross section for lepton pair production from $J/\psi$ decay,
\begin{eqnarray}
  && \!\!\!\!\!\!\!\!\!\!\!\!
   \frac{d \sigma}{d^2 p_{1\perp} d^2 p_{2\perp} dy_1 dy_2 d^2 \tilde b_{\perp} }  =\frac{{\cal C} }{2(2\pi)^7}  \frac{24e^4 e_q^2 }{(Q^2-M^2)^2+M^2 \Gamma^2} \frac{|\phi(0)|^2}{M}
    \nonumber \\ &&\!\!\!\!\!\!\!\!\times
    \int d^2 \Delta_\perp d^2k_\perp d^2 k_\perp'
    \delta^2(k_\perp+\Delta_\perp-q_\perp) \left [ \hat k_\perp'\cdot \hat k_\perp-
     \frac{4( P_\perp \! \cdot \hat k_\perp )( P_\perp\! \cdot \hat k_\perp' )}{M^2} \right ]
     \nonumber \\ && \!\!\!\!\!\!\!\!\times \!
    \left \{ \int \!\! d^2   b_\perp e^{i \tilde b_\perp \cdot
    (k_\perp'\!\!-k_\perp)} \left [ T_A(b_\perp) {\cal
    A}_{in}(x_2,\Delta_\perp) {\cal A}_{in}^*(x_2,\Delta_\perp') {\cal
    F}(x_1,k_\perp){\cal F}(x_1,k_\perp')\!
      + \!( A \!\leftrightarrow \! B)
    \right ] \right .\
     \nonumber \\ &&
    + \!\!\left [  e^{i \tilde b_\perp \cdot (k_\perp'\!\!-k_\perp)}
    {\cal A}_{co}(x_2,\Delta_\perp) {\cal A}_{co}^*(x_2, \Delta_\perp')
    {\cal F}(x_1,k_\perp){\cal F}(x_1,k_\perp')
     \right ]
      \nonumber \\ &&
    + \!\!\left [  e^{i \tilde b_\perp \cdot (\Delta_\perp'\!\!-\Delta_\perp)}
    {\cal A}_{co}(x_1,\Delta_\perp) {\cal A}_{co}^*(x_1, \Delta_\perp')
    {\cal F}(x_2,k_\perp){\cal F}(x_2,k_\perp')
     \right ]
    \nonumber \\  &&+ \!\!
      \left [ e^{i \tilde b_\perp \cdot (\Delta_\perp'\!-k_\perp)}
     {\cal A}_{co}(x_2,\Delta_\perp) {\cal A}_{co}^*(x_1, \Delta_\perp'){\cal F}(x_1,k_\perp){\cal F}(x_2,k_\perp')
     \right ]
        \nonumber \\  &&+ \!\!\!\!
     \left .\
     \left [ e^{i \tilde b_\perp \cdot (k_\perp'\!-\Delta_\perp)}
     {\cal A}_{co}(x_1,\Delta_\perp) {\cal A}_{co}^*(x_2, \Delta_\perp'){\cal F}(x_2,k_\perp){\cal F}(x_1,k_\perp')
     \right ]
      \right \}\!
       \label{fcs}
     \end{eqnarray}
  where $y_1$ and $y_2$ are the final state pions' rapidities, $k_\perp$, $\Delta_\perp$, $k_\perp'$ and $\Delta_\perp'$ are the incoming photon's transverse momenta and the nucleus recoil transverse momenta in the amplitude and the conjugate amplitude respectively.  $\tilde b_\perp$ denotes the transverse distance between the center of the two colliding nuclei. The unit transverse vectors are defined following  the pattern as $\hat k_\perp=k_\perp/|k_\perp|$ and $\hat P_\perp=P_\perp/|P_\perp|$.   A prefactor ${\cal C}$ is introduced here to account for the real part of the amplitude as well as the skewness effect. In our numerical estimations, this coefficient is fixed to be ${\cal C}=1.5$ for RHIC energy, ${\cal C}=1.4$ for LHC energy, and ${\cal C}=1.2$ for EIC energy, following the prescription described in Ref.~\cite{Watt:2007nr}. The longitudinal momentum fraction transferred to the vector meson via the dipole-nucleus interaction is given by $x_g= \sqrt{\frac{P_\perp^2+m^2}{S}}(e^{-y_1}+e^{-y_2})$ with $m$ being lepton mass. ${\cal F}(x,k_\perp)$ describes the probability amplitude for finding a photon that carries a certain momentum with the longitudinal momentum fraction being constrained by $x= \sqrt{\frac{P_\perp^2+m^2}{S}}(e^{y_1}+e^{y_2})$.  The squared  ${\cal F}(x,k_\perp)$ is simply the standard photon TMD distribution $f(x,k_\perp)$. Note that the incoming photon carries different transverse momenta in the amplitude and the conjugate amplitude since we fixed $\tilde b_\perp$~\cite{Vidovic:1992ik,Klein:2018fmp,Zha:2018tlq,Klein:2020jom,Klusek-Gawenda:2020eja,Wu:2021ril,Wang:2021kxm}.

Let us now turn to discuss the final state soft photon radiation effect. Since the emitted soft photon tends to be aligned with the outgoing electron or positron (from the decay of the $J/\psi$), the total transverse momentum of the lepton pair acquired from the recoil effect therefore also points toward the individual lepton's direction, on average.  This naturally generates positive $\cos(2\phi)$ and $\cos 4\phi$ asymmetries of purely perturbative origin for the dilepton system.  The corresponding physics from such final state photon radiation is captured by the soft factor which enters the cross section formula via, 
  \begin{eqnarray}
    \frac{d \sigma(q_\perp)}{d {\cal P.S.}}=\int d^2 q_\perp' \frac{d \sigma_0(q_\perp')}{d {\cal P.S.}}
    S(q_\perp-q_\perp')
  \end{eqnarray}
where $\sigma_0 $ is the leading order Born cross section and $d {\cal P.S.}$ stands for the phase space factor. The soft factor is expanded at the leading order as in~\cite{Hatta:2020bgy,Hatta:2021jcd}, 
\begin{eqnarray}
S(l_{ \perp})\!=\! \delta(l_{ \perp})\!+\! \frac{\alpha_e }{\pi^2 l_{ \perp}^2} \left \{ c_0\!+\!2 c_2 \cos 2\phi\!+\!2 c_4\cos 4\phi+... \right \} 
\label{inte}
\end{eqnarray}
where $\phi$ is the angle between $P_\perp$ and the soft photon transverse momentum $-l_{ \perp}$.  The coefficients can be computed with,
\begin{eqnarray}
c_n= \frac{1}{2\pi}\! \int_0^{2\pi}\!\!\! d\phi \cos (n\phi) \int_{-\infty}^{\infty} \! dy_\gamma 
\frac{1+\cosh(\Delta y_{12})}{2[A \cosh(y_1-y_\gamma)-\cos \phi ][ \cosh(y_2-y_\gamma)+\cos (\phi)]}
\end{eqnarray}
where $A=\sqrt{1+\frac{m^2}{P_\perp^2}}$ and $\Delta y=y_1-y_2$.  When the final state particle mass is much smaller than $P_\perp$, there exits an analytical expressions for the coefficients up to the power correction of $\frac{m^2}{P_\perp^2}$. When $y_1=y_2$, one has $c_0\approx \ln \frac{M^2}{m^2}$, $c_2\approx \ln \frac{M^2}{m^2}-4 \ln 2$ and $c_4\approx \ln \frac{M^2}{m^2}-4 $.  The rapidity dependence of these coefficients are quite mild for RHIC and LHC kinematics and thus are neglected. 

Following the standard procedure, the soft factor in Eq.~\ref{inte} can be extended to all orders by exponentiating the azimuthal independent part to the Sudakov form factor in the transverse position space. The resummed cross section takes the form~\cite{Catani:2014qha,Catani:2017tuc,Hatta:2020bgy,Hatta:2021jcd},
  \begin{eqnarray}
  \frac{d\sigma(q_\perp)}{d {\cal P.S.} }\!&=&\!\! \!\int \!
  \frac{d^2 r_\perp}{(2\pi)^2}
  \left [1\!-\!\frac{2\alpha_e c_2 }{\pi} \cos 2\phi_r +\frac{\alpha_e c_4}{\pi} \cos 4\phi_r\right ]  e^{i r_\perp \cdot q_\perp} e^{- Sud(r_\perp)} \!\! \int \!\! d^2 q_\perp'
  e^{i r_\perp \cdot q_\perp'}  \frac{d\sigma(q_\perp')}{d {\cal P.S.} }.~~~
  \end{eqnarray}
Here $\phi_r$ is the angle between $r_\perp$ and $P_\perp$. The Sudakov factor at one loop is given by~\cite{Hatta:2020bgy,Hatta:2021jcd},
  \begin{eqnarray}
  Sud(r_\perp)=
\frac{\alpha_e}{\pi} {\rm ln} \frac{M^2}{m^2}  {\rm ln}\frac{P_\perp^2}{\mu_r^2}   
  \end{eqnarray}
with $\mu_r=2 e^{-\gamma_E}/|r_\perp|$.
The Sudakov factor plays a critical role in yielding a perturbative source of the high $q_\perp$ tail of lepton pair produced in UPCs~\cite{Klein:2018fmp,Xiao:2020ddm}.

\section{numerical estimations}
We now introduce models/parametrizations used in our numerical calculations. Let us first specify the dipole-nucleus scattering amplitude, which is expressed in terms of dipole-nucleon scattering amplitude ${\cal N}(r_\perp)$~\cite{Kowalski:2003hm,Kowalski:2006hc,Rezaeian:2012ji,Kowalski:2008sa,Kowalski:2007rw},
 \begin{eqnarray}
  N(b_\perp, r_\perp) \approx 1-\left [1-
2\pi B_p  T_A(b_\perp) {\cal N}(r_\perp)\right ]^A
  \end{eqnarray}
where  $B_p=4 {\text GeV}^{-1}$. 
The  dipole-nucleon scattering amplitude is parametrized as~\cite{Bartels:2002cj,Lappi:2010dd,Rezaeian:2012ji,Kowalski:2008sa,Kowalski:2007rw},
\begin{eqnarray}
{\cal N}(r_\perp)=\left \{ 1-\exp \left[-r_\perp^2 G(x_g,r_\perp) \right ] \right \}
\end{eqnarray}
Here $G$ is proportional to the DGLAP evolved gluon distribution in the Bartels, Golec-Biernat and Kowalski
(BGBK) parametrization~\cite{Bartels:2002cj},
\begin{eqnarray}
G(x_g,r_\perp)= \frac{1}{2\pi B_p}\frac{\pi^2}{2N_c} \alpha_s \left ( \mu_0^2+\frac{C}{r_\perp^2} \right )
x f_g \left (x_g, \mu_0^2+ \frac{C}{r_\perp^2} \right )
\end{eqnarray}
with $C$ chosen as $4$ and $\mu_0^2=1.17 \text{GeV}^2$ resulting from the fit~\cite{Kowalski:2006hc} that describes the HERA data quite well.

\begin{figure}[htpb]
  \includegraphics[angle=0,scale=0.38]{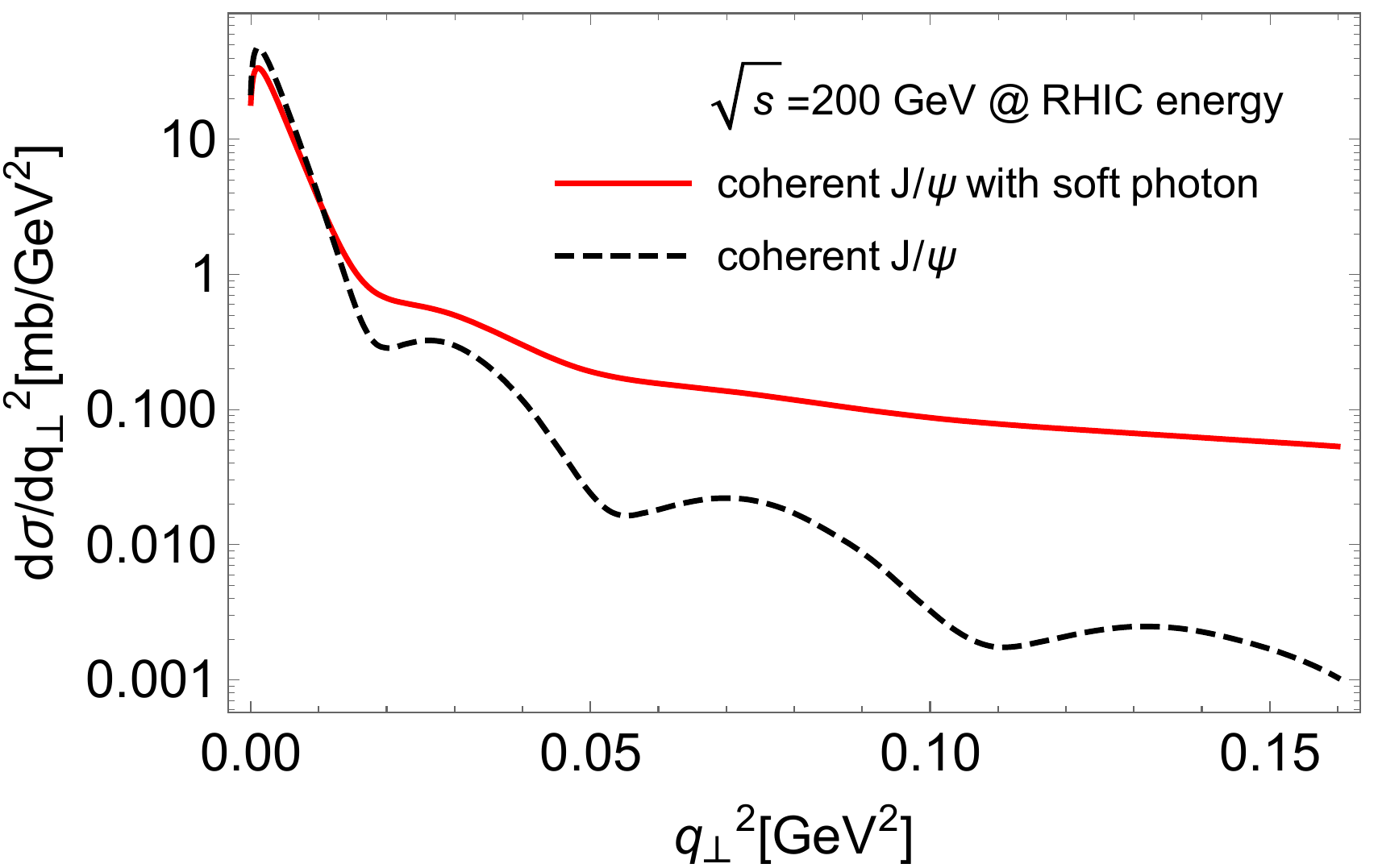}
   \includegraphics[angle=0,scale=0.37]{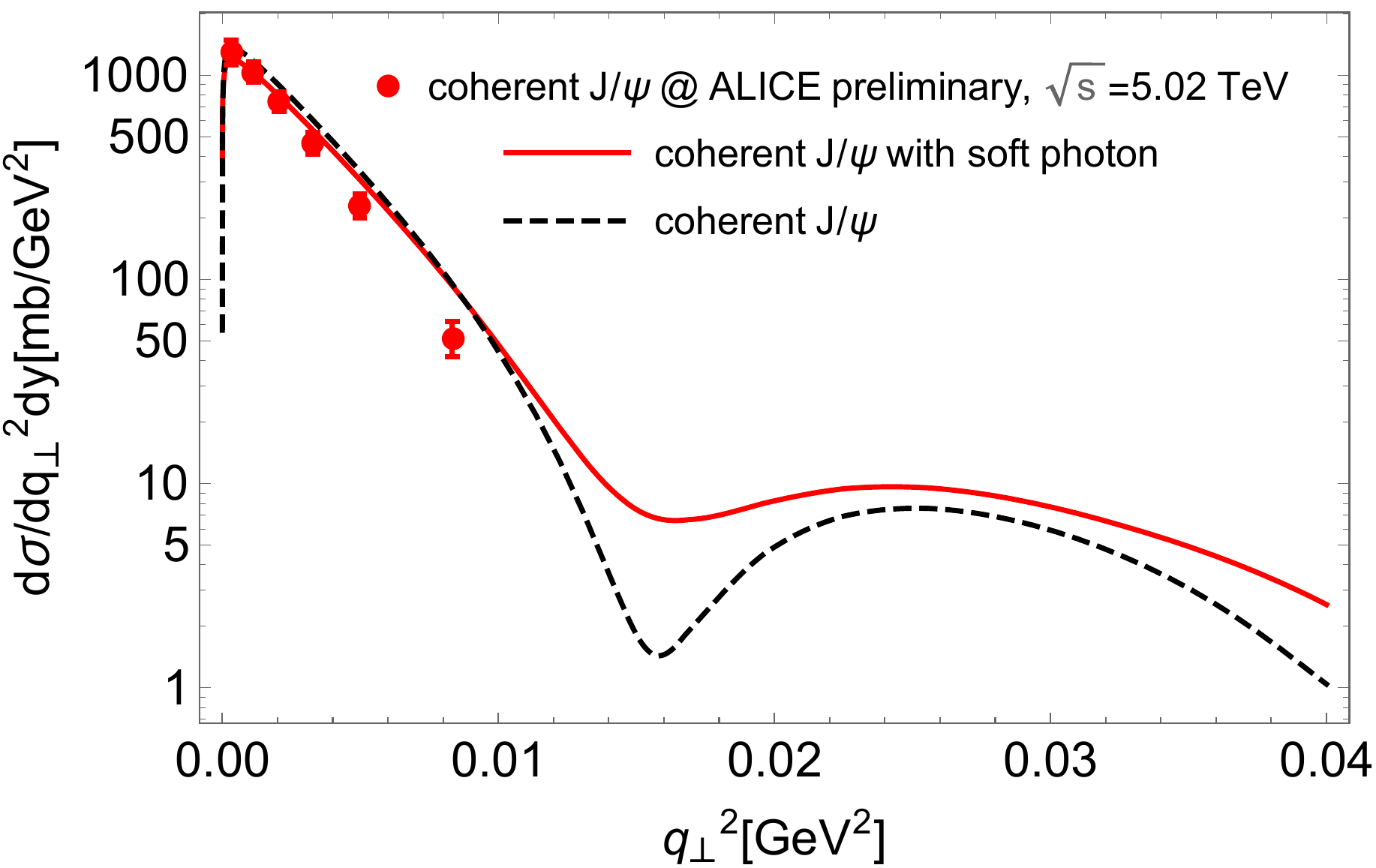}
  \caption{Azimuthal averaged cross section of coherent $J/\psi$ production in unrestricted UPCs at RHIC energy and LHC energy. The rapidity of J/$\psi$ is integrated over the range [-1, 1] for RHIC kinematics and [-0.8, 0.8] for LHC kinematics.} \label{average cs}
\end{figure}

\begin{figure}[htpb]
  \includegraphics[angle=0,scale=0.36]{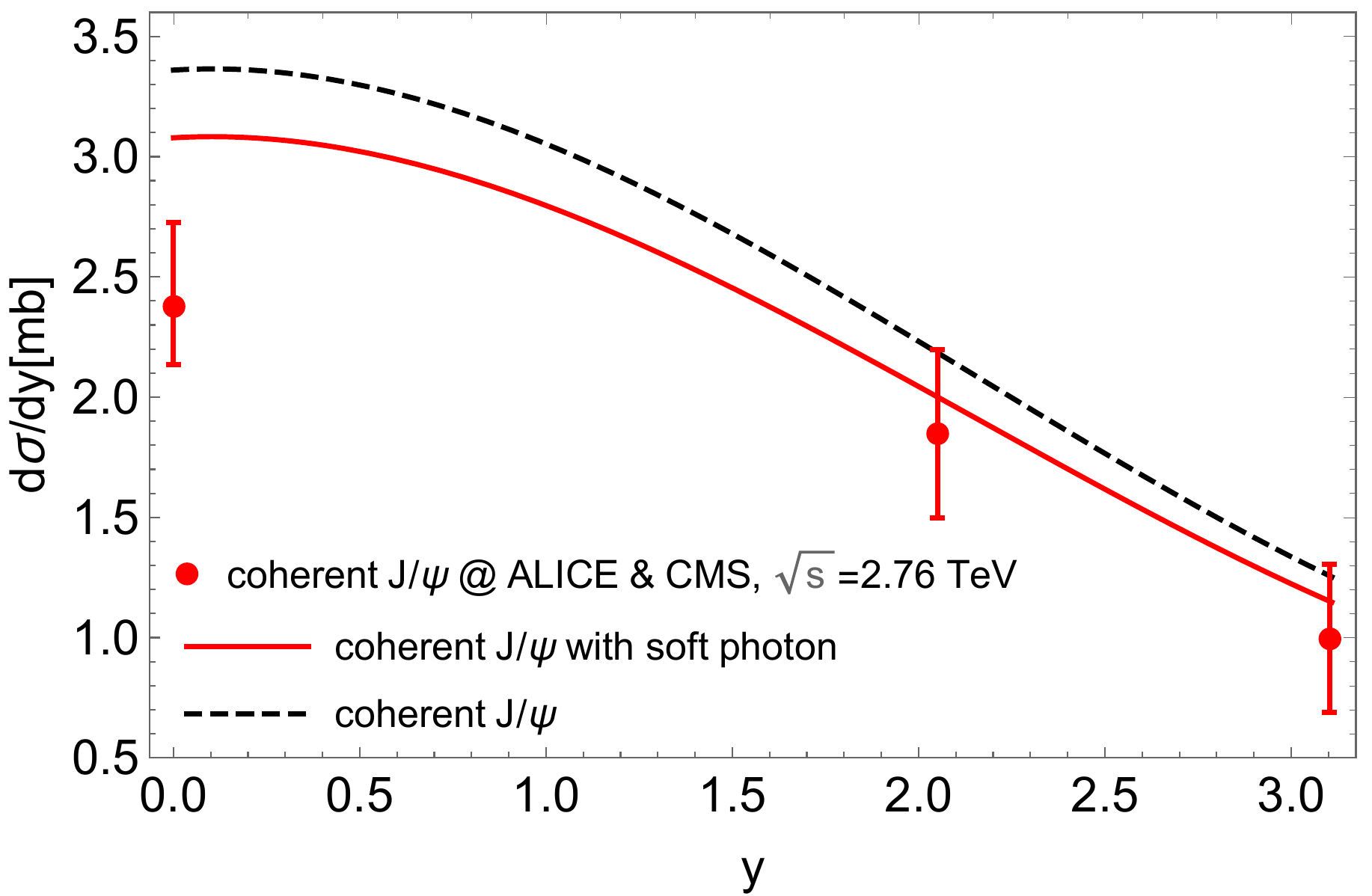}
  \includegraphics[angle=0,scale=0.35]{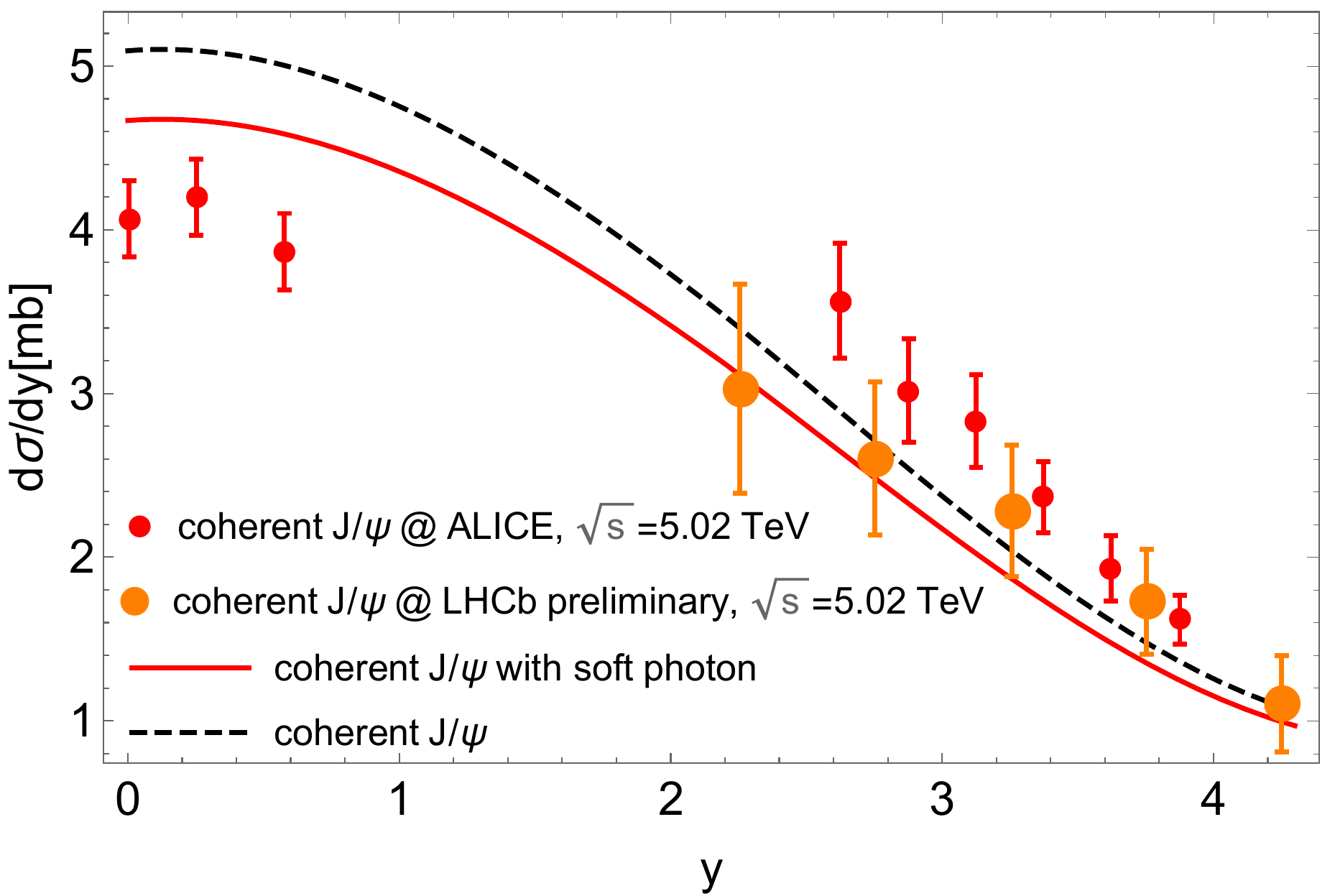}
  \caption{Azimuthal averaged cross section of coherent $J/\psi$ production in unrestricted UPCs at LHC energy. The transverse momentum of J/$\psi$ is integrated over the range [0, 0.2] GeV.} \label{average asy}
\end{figure}

\begin{figure}[htpb]
    \includegraphics[angle=0,scale=0.36]{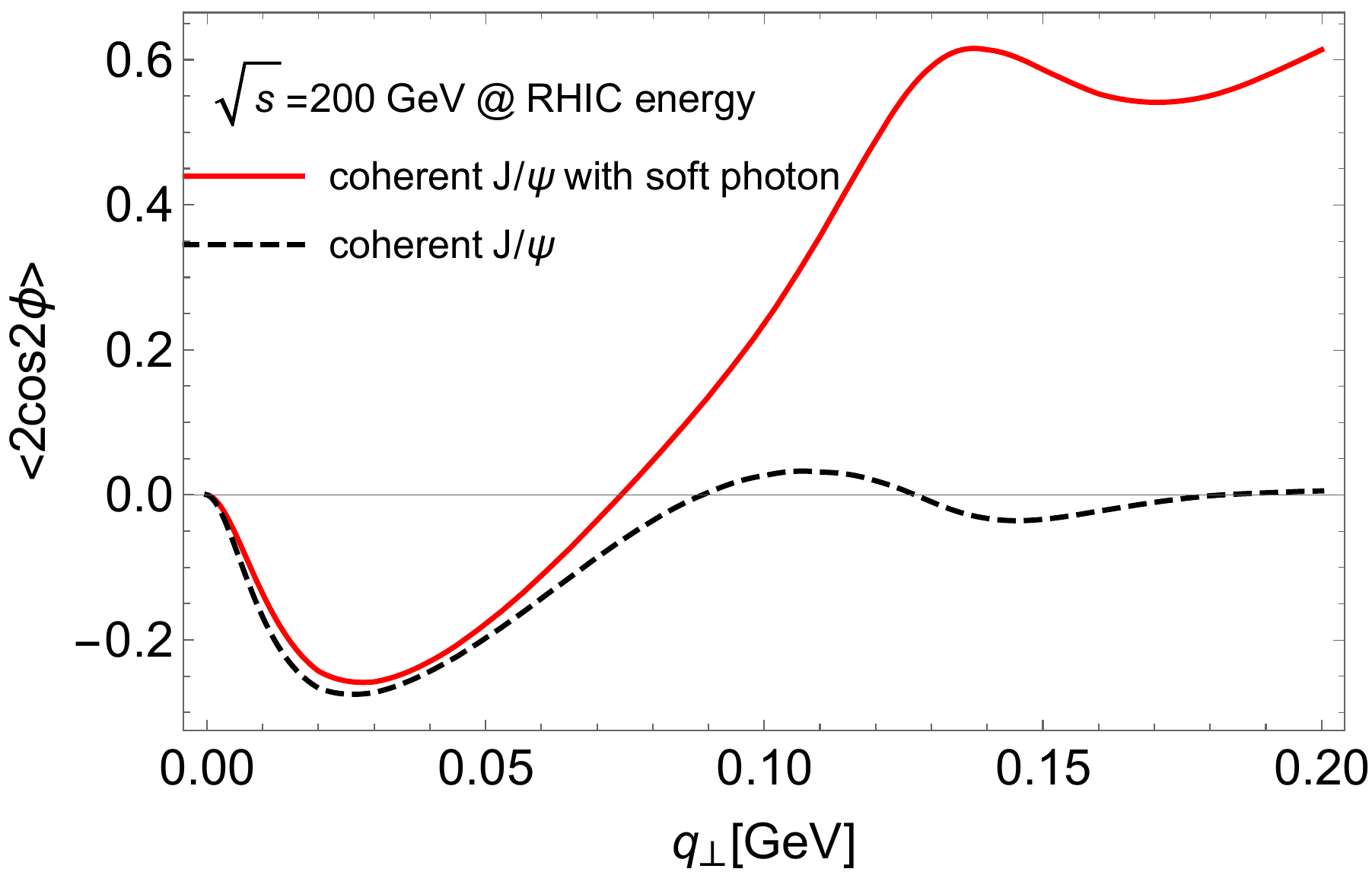}
    \includegraphics[angle=0,scale=0.35]{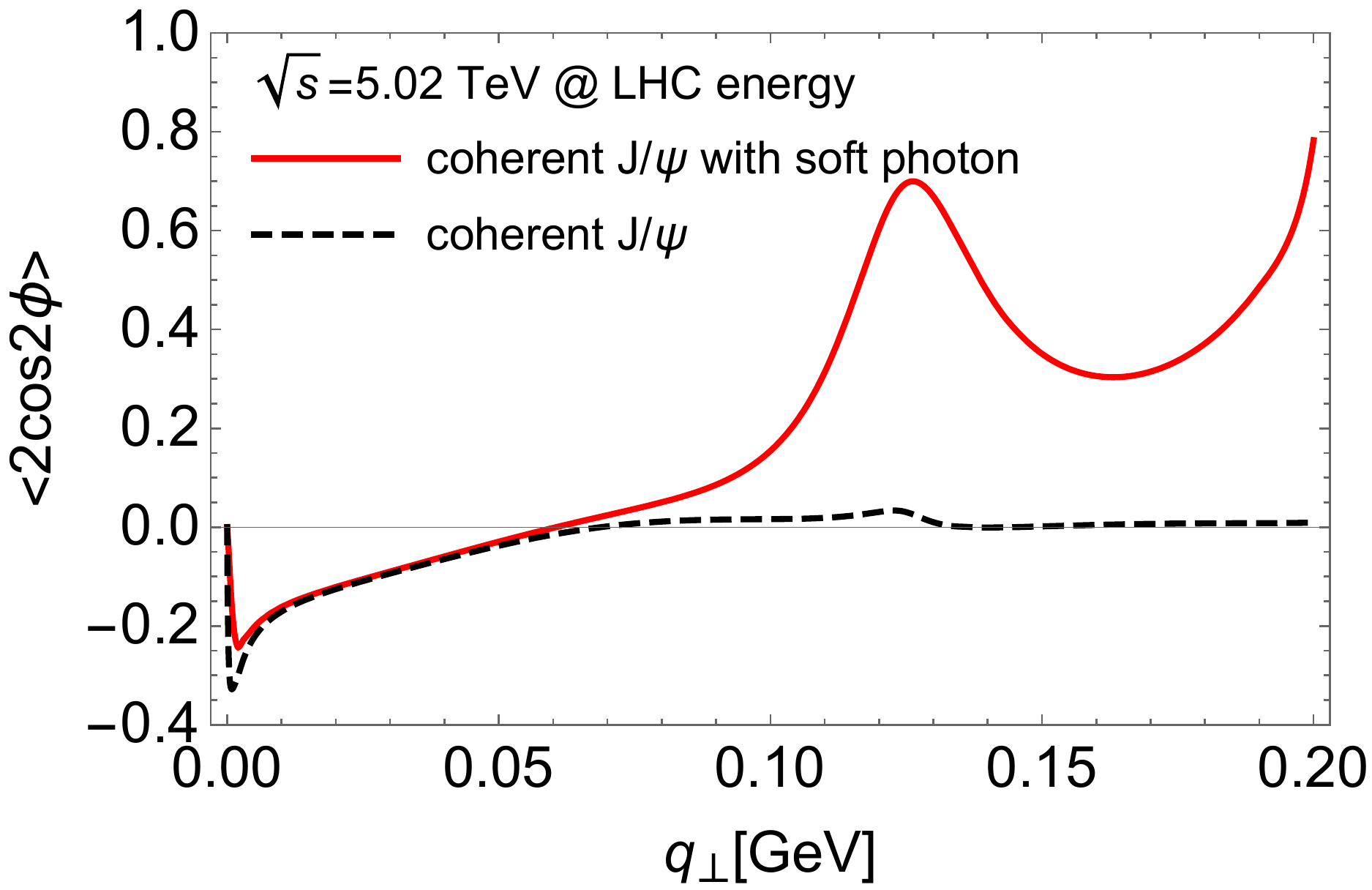}
  \caption{ $\cos 2\phi$ azimuthal asymmetry in coherent $J/\psi$ production at RHIC energy and LHC energy. The rapidity of the di-lepton pair is integrated over the range [-1, 1] at RHIC kinematics and [-0.8,0.8] at LHC kinematics. $J/\psi$ is reconstructed via the decay mode $J/\psi\rightarrow e^+e^-$ at RHIC and  $J/\psi\rightarrow \mu^+\mu^-$ at LHC, respectively.} \label{cos2phi}
\end{figure}

\begin{figure}[htpb]
  \includegraphics[angle=0,scale=0.41]{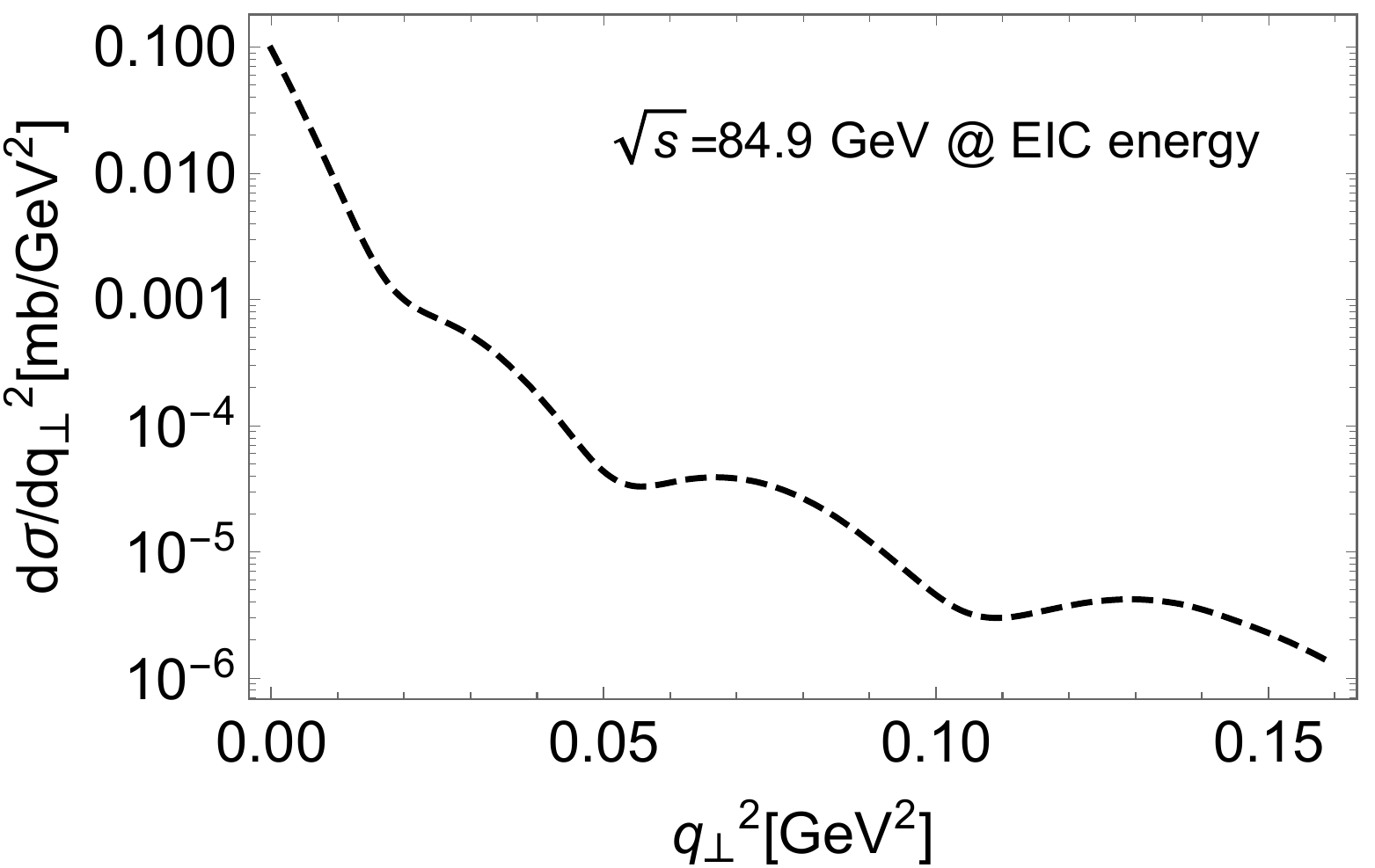}
  \includegraphics[angle=0,scale=0.41]{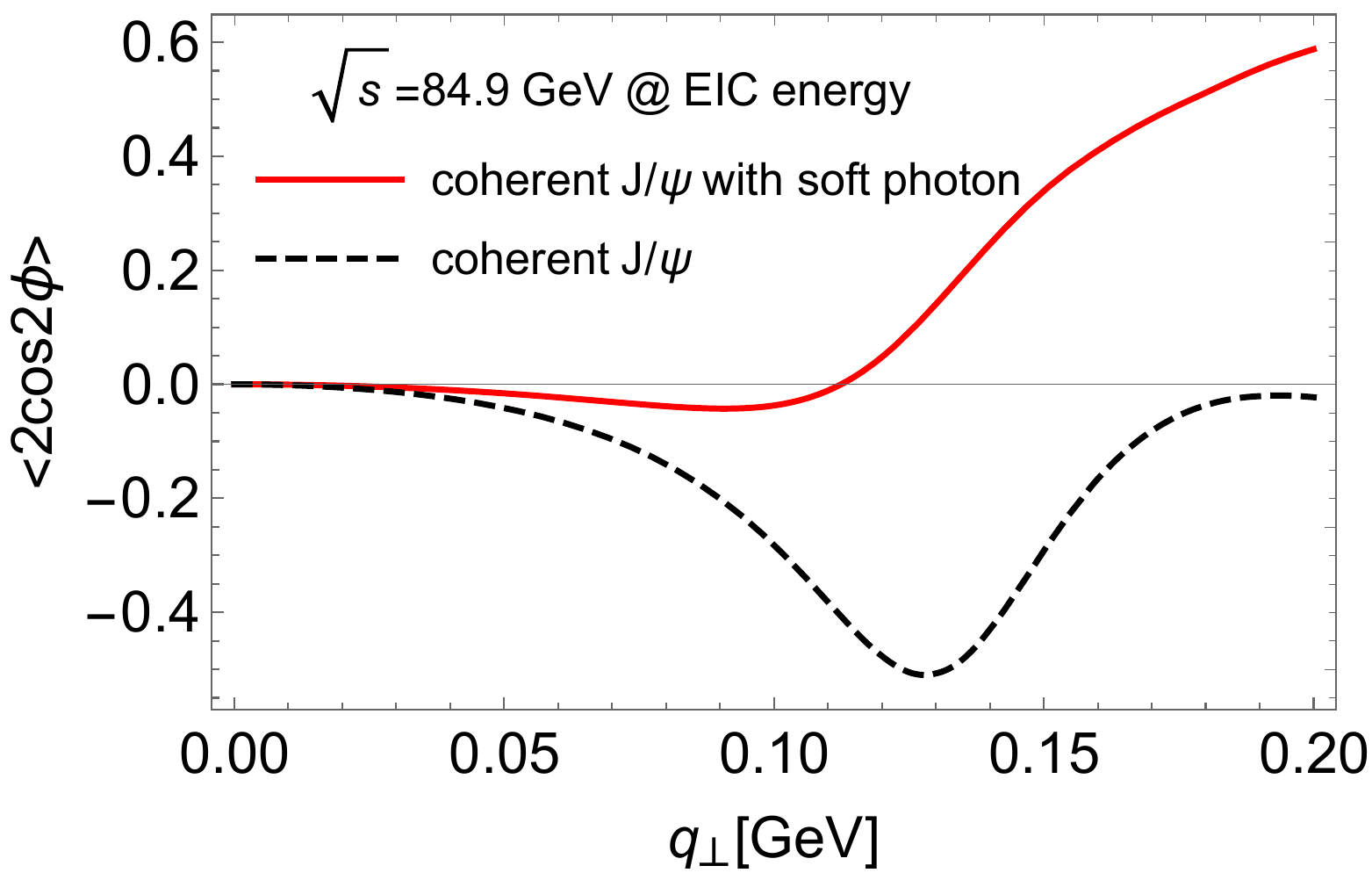}
  \caption{ Azimuthal averaged cross section of coherent $J/\psi$ photoproduction in $e A$ collisions at EIC energy (left panel) and  $\cos 2\phi$ azimuthal asymmetry (right panel) for the same process. The rapidities of $J/\psi$ and its decay product di-electron pair are integrated over the range [2,3] in the Lab frame. The transverse momentum  of  quasi-real photons emitted from the electron is required to be lower than 0.1 GeV.} \label{EIC}
\end{figure}

The nuclear thickness function $T_A(b_\perp)$ is computed with the conventional Woods-Saxon distribution,
 \begin{eqnarray}
 F(\vec k^2)= \int d^3 r e^{i\vec k\cdot \vec r} \frac{C^0}{1+\exp{\left [(r-R_{A})/d\right ]}}
 \end{eqnarray}
where  $R_{A}$(Au: 6.38 fm, Pb: 6.68 fm) is the radius and $d$(Au: 0.535 fm, Pb: 0.546 fm) is the skin depth, and $C^0$ is a normalization factor.  For the scalar part of the vector meson wave function, we use the "Gaus-LC" wave function, also taken from Ref.~\cite{Kowalski:2003hm,Kowalski:2006hc}.
\begin{eqnarray}
\Phi^*(|r_\perp|,z)= \beta z(1-z) \exp \left[-\frac{r_\perp^2}{2R_\perp^2}\right ]
\end{eqnarray}
with $\beta=1.23$, $R_\perp^2=6.5 \ \text{GeV}^{-2}$ for $J/\psi$ meson.  

As for the coherent photon distribution, at low transverse momentum it is commonly computed with the equivalent photon approximation (also often referred to as  the Weizs$\ddot{a}$cker-Williams method) which has been widely used to compute UPC observables(see for example~\cite{Klein:2018fmp,Zha:2018tlq,Klein:2020jom}). In the equivalent photon approximation, $ {\cal F}(x,k_\perp)$ reads,   
  \begin{eqnarray}
    {\cal F}(x,k_\perp)=\frac{Z \sqrt{\alpha_e}}{\pi} |k_\perp|
   \frac{F(k_\perp^2+x^2M_p^2)}{(k_\perp^2+x^2M_p^2)},
  \end{eqnarray}
where $M_p$ is the proton mass. We assume that the charge distribution inside the nucleus is also described by the Woods-Saxon form factor. In the EIC case, the incoming electron serve as the photon source. In this case, we take both the electric charge number $Z$ and form factor $F$ to be 1, and replace $M_p$ with $m_e$ in the denominator to obtain the photon distribution for the electron.

To test the theoretical calculation, We first compute the azimuthal averaged cross section of $J/\psi$ coherent photoproduction and compare them with the experimental measurements at RHIC and LHC for unrestricted UPC events ~\cite{STAR:2021wwq,Schmidke:2021,ALICE:2021tyx}, for which case the impact parameter $\tilde b_\perp$ will be integrated from $2R_A$ to $\infty$. 
As shown in Fig.~\ref{average cs}, our calculation can describe the experimental data quite well, in terms of both the shape and the normalization at low $q_\perp$ for coherent $J/\psi $ production. Here we would like to stress that the perturbative tail generated by the final state soft photon radiation dominates over the primordial distribution determined by the nuclear geometry at large $q_\perp$. This was never pointed out before.  We also plot the unpolarized diffractive $J/\psi$ photoproduction cross section from UPCs at an LHC energy as a function of rapidity. We would like to emphasize that it is crucial to take into account the destructive interference contribution that is enhanced at mid-rapidity, in order to reproduce the observed rapidity dependence of the cross section. Moreover, one notices that it leads to a better agreement with the experimental data in terms of the overall normalization after including the soft photon radiation effect.

The numerical results for the azimuthal asymmetries in coherent $J/\psi$ photoproduction at RHIC and LHC energies
are presented in Fig.~\ref{cos2phi}, where the azimuthal asymmetry, i.e., the average value of $\cos 2 \phi$ is
defined as,
\begin{eqnarray}
\langle \cos( 2\phi) \rangle =\frac{ \int \frac{d \sigma}{d {\cal P.S.}} \cos 2 \phi \ d {\cal P.S.} }
{\int \frac{d \sigma}{d {\cal P.S.}}  d {\cal P.S.}}
\end{eqnarray}
At low $q_\perp$, the asymmetries mainly results from the linear polarization of coherent photons, whereas the asymmetries is overwhelmingly generated by soft photon radiation at relatively large $q_\perp$.  One can see that the asymmetry for $J/\psi$  flips sign as compared to $\rho^0$ production case at low $q_\perp$~\cite{Xing:2020hwh,Zha:2020cst}, mainly due to the fact that the decay product of  $J/\psi$ are spin 1/2 particles, while the decay product of $\rho^0$ are scalar particles.  However, two approaches developed in Ref.~\cite{Xing:2020hwh,Zha:2020cst} predicate quite different the size of the asymmetry at the second peak, though  they yield more or less the same first peak. The origin of this discrepancy remains unknown, and certainly deserves further study in the future.

Our predictions for coherent $J/\psi$ photoproduction in electron-gold nucleus collisions at EIC energy are shown in Fig.~\ref{EIC}. The rapidities are defined in the lab frame, $x= \frac{\sqrt{P_\perp^2+m^2}}{2E_e}(e^{y_1}+e^{y_2})$ and $x_g= \frac{\sqrt{P_\perp^2+m^2}}{2E_A}(e^{-y_1}+e^{-y_2})$, where electron beam and heavy-ion beam energies are 18 GeV and 100 GeV respectively. It is worthwhile to emphasize that the double-slit interference effect is absent in this case. The terms in the last three lines in Eq.\ref{fcs} do not contribute to $J/\psi$  production in eA collisions.  The impact parameter  $\tilde{b}_{\perp}$ is integrated over the range $[0,\infty)$ when computing the coherent cross section in eA collisions. Due to the lack of the double-slit interference effect, the $q_\perp$ shape of the asymmetry is significantly different from that in UPCs.  It will be very interesting to test this theoretical predication at the future EIC.

\section{conclusion}
We have studied coherent $J/\psi$ photoproduction in UPCs and in eA collisions using the dipole model with all parameters fitted to HERA data. Our calculations are in good agreement with the experimental measurements performed at low transverse momentum from RHIC and LHC. It has been demonstrated that double-slit interference effect and final state soft photon effect are the absolutely crucial ingredients to correctly account for the $t$ and $y$ dependent shape, as well as the overall normalization of the coherent $J/\psi$ photoproduction in UPCs.  We further computed the azimuthal asymmetries arising from the linear polarization of the incident photons and the final state soft photon radiation for $J/\psi$ production in UPCs and in eA collisions.  As these polarization dependent observables are sensitive to nuclear geometry, they may provide complementary information on the gluon tomography of nucleus at small $x$. On top of this, the double-slit interference effect deserves to be studied in more details in high energy scatterings in its own right.

\begin{acknowledgments}
  J. Zhou has been supported by the National Science Foundations of China under Grant No.\ 12175118. Y. Zhou has been supported by the Natural Science Foundation of Shandong Province under Grant No. ZR2020MA098. C. Zhang has been supported by the National Science Foundations of China under Grant No.\ 12147125. This work is supported in part by the U.S. DOE Office of Science under contract Nos. DE-SC0012704, DE-FG02-10ER41666, and DE-AC02-98CH10886. 
\end{acknowledgments}

\bibliography{ref.bib}

\end {document}